\documentclass[aip,amsmath,amssymb,reprint]{revtex4-1}

\draft 

\usepackage{graphicx}
\usepackage{dcolumn}
\usepackage{bm}

\usepackage{dsfont}
\usepackage[usenames,dvipsnames]{xcolor}
\usepackage{pstricks}
\usepackage[tight]{subfigure}
\usepackage{verbatim}
\usepackage{units}
\usepackage{multirow}
\usepackage{enumitem}
\usepackage{mathrsfs}
\usepackage{leftidx}
\usepackage{xspace}
\usepackage{diffcoeff}  
\usepackage{bbm}
\usepackage{amsfonts,amssymb,amsmath}
\usepackage{mathtools}

\usepackage{array} 
\usepackage{tabularx}
\usepackage[normalem]{ulem}
\usepackage{cancel}
\usepackage{bbold}
\usepackage{pifont}
\usepackage{hyperref}
\hypersetup{colorlinks=true,linktoc=all,linkcolor=blue,breaklinks=true,citecolor=blue,urlcolor=blue}

\usepackage[utf8]{inputenc}
\usepackage[T1]{fontenc}
\usepackage{mathptmx}
\usepackage{etoolbox}
\DeclareSymbolFont{newfont}{OML}{cmm}{m}{it}
\DeclareMathSymbol{\epsilon}{3}{newfont}{15}
\DeclareMathAlphabet{\mathcal}{OMS}{cmsy}{m}{n}

\makeatletter
\def\@email#1#2{%
 \endgroup
 \patchcmd{\titleblock@produce}
  {\frontmatter@RRAPformat}
  {\frontmatter@RRAPformat{\produce@RRAP{*#1\href{mailto:#2}{#2}}}\frontmatter@RRAPformat}
  {}{}
}%
\makeatother

\newcommand{\beq}{\begin{equation}}
\newcommand{\eeq}{\end{equation}}
\newcommand{\rme}{\text{e}}

\newcommand*{\Dmu}{\Delta\mu}

\DeclarePairedDelimiterX\braket[2]{\langle}{\rangle}{#1\,\delimsize\vert\,\mathopen{}#2}
\DeclarePairedDelimiterX\ketbra[2]{\lvert}{\rvert}{#1\,\delimsize\rangle\mathopen{}\delimsize\langle\,\mathopen{}#2}

\DeclarePairedDelimiterX\Braket[2]{(}{)}{#1\,\delimsize\vert\,\mathopen{}#2}
\DeclarePairedDelimiterX\Ketbra[2]{\lvert}{\rvert}{#1\,\delimsize)\mathopen{}\delimsize(\,\mathopen{}#2}

\newcommand{\kB}{k_\text{B}}
\newcommand{\kBT}{k_\text{B}T}

\begin{document}


\title{Thermal junctions controlled with Aharonov-Bohm phases} 



\author{Jos\'e Balduque}
\affiliation{Departamento de F\'isica Te\'orica de la Materia Condensada, Universidad Aut\'onoma de Madrid, 28049 Madrid, Spain\looseness=-1}
\affiliation{Condensed Matter Physics Center (IFIMAC), Universidad Aut\'onoma de Madrid, 28049 Madrid, Spain\looseness=-1}
\author{Adri\'an Mecha}
\affiliation{Departamento de F\'isica Te\'orica de la Materia Condensada, Universidad Aut\'onoma de Madrid, 28049 Madrid, Spain\looseness=-1}
\author{Rafael S\'anchez}
\affiliation{Departamento de F\'isica Te\'orica de la Materia Condensada, Universidad Aut\'onoma de Madrid, 28049 Madrid, Spain\looseness=-1}
\affiliation{Condensed Matter Physics Center (IFIMAC), Universidad Aut\'onoma de Madrid, 28049 Madrid, Spain\looseness=-1}
\affiliation{Instituto Nicol\'as Cabrera, Universidad Aut\'onoma de Madrid, 28049 Madrid, Spain\looseness=-1}


\date{\today}

\begin{abstract}
Unlike charge, heat flows are difficult to control.
We show that, in mesoscopic conductors, electronic thermal currents can be manipulated with a magnetic field by using the Aharonov-Bohm effect: the magnetic control of the interference pattern enhances the thermoelectric effect, while heat transport can be totally suppressed. In a three-terminal configuration, the flux-induced broken reciprocity generates a non-local thermoelectric response and translates to the circulation of heat. This way, efficient thermoelectric generators, thermal switches and thermal circulators, as well as energy harvesters can be defined for minimally disturbing thermal management at the nanoscale. 
\end{abstract}


\maketitle 


\section{Introduction}
\label{sec:intro}

Transport through mesoscopic conductors is determined by the phases that electrons accumulate in the different trajectories between terminals~\cite{heikkila_book}. In systems with several barriers~\cite{buttiker:1988}, or in molecular junctions~\cite{cuevas_book}, the interference of kinetic phases acquired during multiple reflections between two scattering regions result in Fabry-P\'erot or Fano resonances. Trajectories enclosing a magnetic field acquire an additional phase which leads to the Aharonov-Bohm effect~\cite{aharonovbohm}. If the trajectories are confined in a quasi-one-dimensional ring, the current presents flux-dependent oscillations~\cite{gefen_quantum_1984,buttiker:1984} that can be controlled experimentally~\cite{webb_observation_1985,datta_novel_1985} and revealed the nonlocal nature of quantum transport~\cite{buttiker_four_1986,benoit_asymmetry_1986}. They are a useful tool to measure magnetic asymmetries of the transport coefficients~\cite{Casimir1945,buttiker_symmetry_1988,butcher:1990,jacquod_onsager_2012} and fluctuations~\cite{nakamura_nonequilibrium_2010,nakamura_fluctuation_2011}, measurement-induced dephasing~\cite{aleiner_dephasing_1997,buks_dephasing_1998}, scattering phases~\cite{yacobi_coherence_1995,yeyati_aharonov_1995,hackenbroich_transmission_1996,oreg_electron_1997,bruder_aharonov_1996,yacoby_cohernece_1996,schuster_phase_1997,cernicchiaro_channel_1997,kobayashi_tuning_2002,kobayashi_mesoscopic_2003,sigrist_magnetic_2004,kobayashi_fano_2004,avinunKalish_crossover_2005,buchholz_control_2010}, heat transport~\cite{buchholz_noise_2012}, or the coherence in topological states~\cite{behner_aharonov_2023} and fractional excitations in the quantum Hall regime~\cite{biswas_anomalous_2024,kim_aharonov_2024}. The interplay of different such interferences may also arise in quantum dot~\cite{hofstetter_kondo_2001,entin_fano_2002,kubala_aharonov_2003,he_electron_2006,emary_dark_2007,rai_magnetic_2012,bedkihal_flux_2013,danjou_anomalous_2013,lu_optimal_2019} and atomtronic~\cite{lau_atomtronic_2023} structures.

The energy dependence of kinetic phases has motivated coherent conductors to be useful for thermoelectrics~\cite{sivan:1986,paulsson:2003,finch:2009,bergfield:2009,karlstrom:2011,trocha:2012,gomezsilva:2012,hershfield:2013,nakpathomkun:2010,rincon-garcia_molecular_2016,cui_peltier_2017,miao_influence_2018,brun2019,hamill_quantum_2023}, as it naturally breaks electron-hole symmetry in an interferometer~\cite{hofer:2015,Vannucci2015, samuelsson:2017,haack:2019,haack_nonlinear_2021,genevieveqpc,extrinsic,blasi_hybrid_2022}. In three terminal configurations~\cite{entin_three_2010,hotspots,sanchez:2011,entin_three_2012,mazza:2014,mazza:2015}, this property allows for the conversion of heat injected in a junction into electrical power, only relying on the properties of the junction~\cite{extrinsic}, and in the absence of charge injection from the heat source. When the contact is given by a scanning probe~\cite{genevieveqpc,extrinsic,balduque_coherent_2023}, the heat source can be disconnected at will. However, one does not have such possibility in most experimental realizations, where the injection of heat is permanent and difficult to control. 

For the development of nanoscale and quantum technologies, it would be desirable to achieve a similar degree of control of heat currents as for charge to manage the excess heat in such devices.  While the control of charge currents is routinely done by external voltages, such a contact for heat is limited by the Joule and the Seebeck effects~\cite{dubi_colloquium_2011,benenti:2017,zhang_local_2019}. Despite the formidable advances in the last years~\cite{giazotto:2006,riha_heat_2016,pekola_colloquium_2021,majidi_heat_2024}, this goal is still challenging, with few experimental demonstrations of thermal devices (thermal diodes, transistors or circulators) so far~\cite{giazotto_josephson_2012,martinezPerez_rectification_2015,ronzani_tunable_2018,senior_heat_2020,gubaydullin_photonic_2022,subero_bolometric_2023,navarathna_passive_2023,fedorov_nonreciprocity_2024}, especially in normal conductors~\cite{jezouin_quantum_2013,thierschmann_thermal_2015,dutta_thermal_2017}. In particular, junctions that control noninvasively how heat is injected, such as thermal switches~\cite{karimi_coupled_2017,donald,haack:2019,Kirsanov2019}, modulators~\cite{hwang_phase_2024} or circulators~\cite{chiraldiode,hwang:2018,acciai:2021,diaz_qutrit_2021}, would enable the definition of integrated heat circuits~\cite{wang_thermal_2007,segal_nonlinear_2008,li_colloquium_2012}. Single-electron transistors have been proposed~\cite{tsaousidou_thermal_2007,zianni_coulomb_2007,kubala_violation_2008} which are restricted to very low powers.

\begin{figure}[b]

\includegraphics[width=\linewidth]{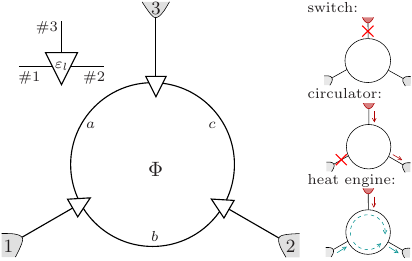}
\caption{\label{fig:scheme}\small
Single-channel three-terminal quantum ring. Each terminal is in local equilibrium with an electrochemical potential $\mu_l$ and a temperature $T_l$ ($l$=1,2,3) and coupled to the ring by a beam-splitter with coupling $\varepsilon_l$ dividing the ring in three segments of length $L_\alpha$ ($\alpha$=a,b,c). A magnetic flux $\Phi$ pierces the ring inducing a magnetic phase $\pm\phi_\alpha$ in the electrons propagating along each segment. Insets show the scheme for the ring-lead junctions and the three main operations considered here. 
}
\end{figure}

Recently Haack {\it et al.} have proposed the Aharonov-Bohm effect 
as a mechanism to control both the thermoelectric response~\cite{guttman_thermopower_1995,blanter_aharonov_1997,lu_thermoelectric_2014} and the flow of heat in two-terminal ring conductors by using the magnetic field as a knob~\cite{haack:2019,haack_nonlinear_2021,blasi_hybrid_2022}. Following these ideas, we investigate a minimal model of a multiterminal ring able to exploit the nonlocal features of the Aharonov-Bohm effect to be used as a thermoelectric engine, a thermal switch and a heat circulator, as illustrated in Fig.~\ref{fig:scheme}, furthermore allowing for an analytical description of the relevant processes in terms of the properties and symmetries of the transport coefficients.  This is done based on three main properties: (i) The interplay of the kinetic and magnetic phases leading to an enhanced thermoelectric effect; (ii) the flux-dependent but energy-independent destructive interference of the Aharonov-Bohm scattering leading to the suppression of all (charge and heat) currents, hence to a heat switch that avoids undesired thermoelectric effects, and (iii) the broken time reversal symmetry leading to asymmetric transport coefficients in multiterminal configurations and hence to thermal circulator and energy harvester~\cite{entin_three_2010,hotspots,thierschmann:2015,roche:2015,hartmann:2015,jordan:2013,jaliel:2019,dorsch:2020,dorsch_characterization_2021,khan_efficient_2021,haldar_microwave_2024} effects. 
We pay special attention to the role of dephasing~\cite{buttiker:1986,buttiker:1988,dejong_semiclassical_1996} in two-terminal configurations and of nonreciprocal transport in three-terminal setups. 
We will focus on switches and circulators in {\it all-thermal} configurations i.e., imposing that no charge but only heat is injected from the different terminals by assuming open-circuit conditions, this way defining contactless onchip thermal devices tuned by flux. For comparison, we will also consider {\it electrothermal} cases which relax that condition by allowing charge to flow and facilitate a simple analytical treatment in limiting cases. 

The remaining of the manuscript is organized as follows. In Sec.~\ref{sec:scattering} we introduce the theoretical description of our model based on scattering theory. Sections \ref{sec:2t} and \ref{sec:3t} present the results for two and three terminal configurations, respectively, with the conclusions being discussed in Sec.~\ref{sec:conclusion}.

\section{Scattering theory}
\label{sec:scattering}

Before describing transport, it is useful to describe the states in the ring. Solving the Schr\"odinger equation for a single mode closed ring of circumference $L$ pierced by a magnetic flux $\Phi$ yields a spectrum given by the eigenenergies~\cite{buttiker_josephson_1983}
\beq
\label{eq:eigenen}
\epsilon_n=U_r+\frac{h^2}{2mL^2}\left(n-\frac{\phi}{2\pi}\right)^2,
\eeq
where $U_r$ is the bottom of the confinement potential, $\phi=2\pi\Phi/\Phi_0$ is the phase accumulated upon performing a loop around the ring, with $\Phi_0=h/c$ being the flux quantum, $h$ the Planck constant, and $m$ the mass of the electron
(see also Ref.~\onlinecite{ihn_semiconductor_2009}). 

When connected to leads, the states of the ring mediate the transmission of electrons through the conductor terminals~\cite{buttiker:1984}.
We will consider here two- and three- terminal configurations. The most general case is sketched in Fig.~\ref{fig:scheme}. It is decomposed in three connected three-channel junctions (one for each terminal and one in each ring segment) treated as beam-splitters. For simplicity, we assume that the coupling of the terminal channels with the ring ones, $\varepsilon_l$, is symmetric in each junction $l$, leading to a scattering matrix~\cite{buttiker:1984}:
\begin{eqnarray}\label{Sal}
\displaystyle
{
S^\triangledown_{\alpha l}=
\left(\begin{array}{ccc}
-\rme^{i2\chi_\alpha}\eta_l^-/2 & \rme^{i\varphi_\alpha^+}\eta_l^+/2 & \rme^{i\varphi_\alpha^+}\sqrt{\varepsilon_l}\\ 
\rme^{i\varphi_\alpha^-}\eta_l^+/2 & -\eta_l^-/2 & \sqrt{\varepsilon_l}\\
\rme^{i\varphi_\alpha^-}\sqrt{\varepsilon_l} & \sqrt{\varepsilon_l} & \eta_l^-{-}1\\
\end{array}  \right)
}
\end{eqnarray}
for each junction, with the elements ordered as in the inset of Fig.~\ref{fig:scheme}. The phases $\varphi_\alpha^\pm=\chi_\alpha\pm\phi_\alpha$ accumulated by the waves in the segment $\alpha\in\{a,b,c\}$  when propagating clock- ($+$) or anticlockwise ($-$) contain a kinetic, $\chi_\alpha=\sqrt{2m(E-U_0-e\xi_\alpha V_{g,\alpha})}L_\alpha/\hbar$, where $U_0$ is the energy of the lowest ring sub-band, and a magnetic part, $\phi_\alpha$, both depending on the length of the segment $L_\alpha$. They can be tuned by means of the voltage $V_{g,\alpha}$ applied to plunger gates, with $\xi_\alpha=C_\alpha/C$ being the gate capacitive lever arm, or of the magnetic phase, $\phi=\sum_\alpha\phi_\alpha$. It is convenient to invoke gauge invariance and accumulate the magnetic phase in a single segment, e.g. $\phi_a=\phi$, $\phi_b=\phi_c=0$. The coupling between terminal $l\in\{1,2,3\}$ and the ring is given by $\varepsilon_l\in[0,1/2]$, which is used to define $\eta_l^\pm=1\pm\sqrt{1-2\varepsilon_l}$. The two-terminal configuration is trivially obtained from this by simply setting $\varepsilon_3=0$ and $L_c=0$. In what follows, and in order to emphasize the asymmetries introduced by the phases, we will consider symmetric configurations, with all nonvanishing couplings and lengths being equal, $\varepsilon_l=\varepsilon$ and $L_\alpha=L/N$, with $N=2,3$ being the number of terminals, and the couplings to be energy independent. Also, lengths will be expressed in units of $l_0=2\pi\hbar/\sqrt{2mk_BT}$, which
for GaAs rings is of the order of 730 nm at $T=\unit[1]{K}$. We assume the simplest model with a single-channel ring which is nevertheless relevant for experimental realizations in two-dimensional electron gases~\cite{vanderwiel_electromagnetic_2003}. The contribution of additional channels of different lengths results in the randomization of the kinetic phase, hence contributing to dephasing, see below.

With these matrices, one obtains the scattering matrix ${\cal S}_{l'l}(E)$ of the whole conductor by identifying the outgoing waves of a junction with the ingoing ones of the next one along the ring. The procedure is detailed in textbooks~\cite{Datta1995}, see also Ref.~\onlinecite{extrinsic}. A relevant quantity for transport will be the difference $\mu-U_r$ set by the equilibrium chemical potential of the terminals and the overall potential $U_r\equiv U_0+eV_g$: it determines the kinetic phase of the electrons scattered at the Fermi energy. We will for simplicity chose $\mu=0$ as a reference energy and modulate $U_r$ with a gate voltage globally acting on the ring in the following results. Different from Ref.~\onlinecite{haack:2019}, we will furthermore ignore asymmetries induced by the gate voltages setting $V_g\equiv V_{g,\alpha}$, and consider only the effect of the ring potential $U_r$.

\subsection{Transport}
\label{sec:transport}

With the scattering matrix of the system, ${\cal S}_{l'l}(E)$, we write the charge and heat currents~\cite{moskalets-book}:
\begin{equation}
\label{eq:currents}
I_l=e\int_{-U_r}^\infty{dE} i_l(E)\quad{\rm and}\quad
J_l=\int_{-U_r}^\infty{dE} (E-\mu_l)i_l(E),
\end{equation}
in terms of the spectral particle currents:
\begin{equation}
\label{eq:spectr}
i_l(E)=\frac{2}{h}\sum_{l'}{ \cal T}_{l'l}(E)[f_l(E)-f_{l'}(E)],
\end{equation}
where ${\cal T}_{l'l}(E)=|{\cal S}_{l'l}(E)|^2$ is the transmission probability for an electron injected from terminal $l$ to be absorbed by terminal $l'$, and $f_l(E)=\{1+\exp[(E{-}\mu_l)/\kBT_l]\}^{-1}$ is the Fermi distribution function of terminal $l$ having an electrochemical potential $\mu_l$ and a temperature $T_l$; the factor 2 accounts for spin degeneracy. Both particle and heat currents are defined as positive when emitted by a reservoir. In the presence of a magnetic field, the scattering matrix becomes asymmetric, with reciprocity being expressed as ${\cal S}_{l'l}(\Phi)={\cal S}_{ll'}(-\Phi)$~\cite{buttiker_symmetry_1988}, a property that affects the inelastic thermopower~\cite{saito_thermopower_2011,sanchez:2011,bedkihal_probe_2013} and that we will exploit in multiterminal configurations.

Scattering theory guarantees charge and energy conservation: $\sum_lI_l=0$ and $\sum_l (J_l+\mu_lI_l)=0$. The second term in the later equation corresponds to power dissipated as Joule heating. For the thermoelectric engine, we define the power output
\beq
P=\Delta \mu I_1,
\eeq
with $\Delta \mu=\mu_2-\mu_1$ so that it is positive when generated from the conversion of heat: electrons flow from lower to higher electrochemical potential terminals.

\subsection{Open-circuit condition}
\label{sec:alltherm}

Occasionally, we will require that some terminals inject heat but no charge, on average, into the conductor. This will be the case of the all-thermal operations and of the heat source of the three-terminal energy harvester. This is done by assuming an open circuit condition, such that the chemical potential of the involved terminals develops to a value $\mu_l$ that maintains a vanishing charge current. We then need to solve the equations $I_l(\mu_l)=0$ for those terminals. 

\subsection{Dephasing}
\label{sec:dephas}

To evaluate the effect of decoherence in scattering theory, one introduces fictitious terminals which absorb and reinject electrons with the same energy but a randomized phase~\cite{buttiker:1986,buttiker:1988,dejong_semiclassical_1996}. In order to do so without introducing additional interference, the dephasing probe terminals need two channels coupled to the conductor via scattering matrices of the type~\cite{buttiker:1986}
\begin{eqnarray}\label{Sdeph}
\displaystyle
{
S_\alpha^\Diamond=
\left(\begin{array}{cccc}
0 & \sqrt{1{-}\lambda_\alpha} & \sqrt{\lambda_\alpha} & 0\\ 
\sqrt{1{-}\lambda_\alpha} & 0 & 0 & \sqrt{\lambda_\alpha}\\
\sqrt{\lambda_\alpha} & 0 & 0 & -\sqrt{1{-}\lambda_\alpha}\\
0 & \sqrt{\lambda_\alpha} & -\sqrt{1{-}\lambda_\alpha} & 0\\
\end{array}  \right),
}
\end{eqnarray}
where $\lambda_\alpha \in [0,1]$ parametrizes the coupling of the probe to segment $\alpha$ of the ring.
We are avoiding here the introduction of additional (and unnecessary) phases.
To satisfy that the spectral current $i_\alpha(E)$ vanishes, i.e.:
\beq
\label{eq:deph_cond}
\sum_{l} {\cal T}_{l\alpha}(E)[\tilde{f}_\alpha(E)-f_{l}(E)]=0,
\eeq
where $l$ includes all other terminals (fictitious or not), one needs to impose the condition that the probe has a nonthermal distribution $\tilde f_\alpha(E)$ that solves Eq. \ref{eq:deph_cond} for each energy. Therefore, the probes do not affect the conservation of particle and energy spectral currents in the conductor, and also $I_\alpha=J_\alpha=0$. In the two-terminal setup, we will restrict to the case where each segment is coupled to a different dephasing probe with equal strength $\lambda_\alpha=\lambda$. In the three-terminal setup, we will for simplicity assume that $\lambda_a=\lambda_c=0$ and only $\lambda_b=\lambda$ can be finite. Note that the presence of multiple channels in the ring can be another source of momentum indetermination.

\section{Two terminals}
\label{sec:2t}
\begin{figure}[t]
    \includegraphics[width=\linewidth]{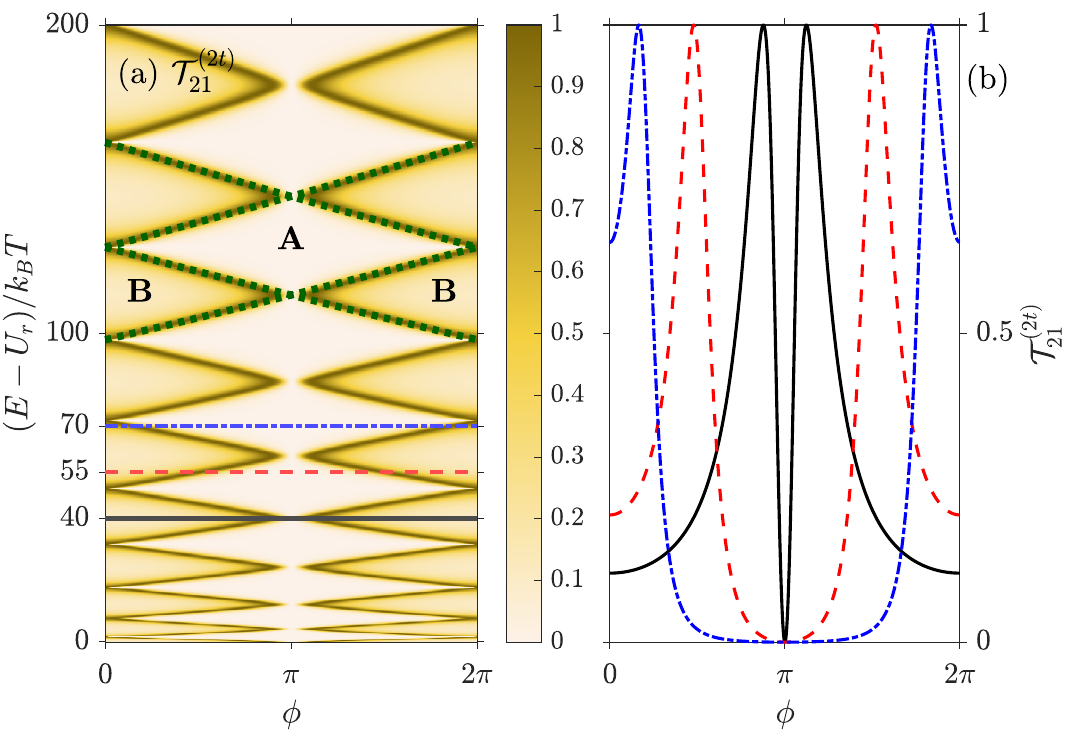}

    \caption{\label{fig:2ttransmiss}\small Two-terminal transmission probability (a) as a function of energy and the magnetic phase, for a ring of circumference $L=l_0/2$ with  $\varepsilon=0.25$. Dotted green lines mark the energy of the closed ring eigenvalues $\epsilon_n$ for $n=-4,\pm5,6$. (b) Cuts of the previous at the energies marked by the lines with the corresponding linestyles: $E-U_r=40\kBT, 55\kBT,70\kBT$. }
\end{figure}
We start with the two-terminal configuration, which has been shown to work as a thermoelectric engine and switch~\cite{haack:2019,haack_nonlinear_2021}. 
In this configuration, we set $\varepsilon_3=0$ and $L_c=0$ in the scheme of Fig.~\ref{fig:scheme}.
In the transmission probability through the quantum ring in that configuration and in the absence of dephasing ($\lambda=0)$~\cite{buttiker:1984,haack:2019}
\begin{align}
    \label{eq:T_two_terminals}
    {\cal T}_{12}^{(2t)}(E)=
     \frac{8\varepsilon^2\sin^2{\chi}(1+\cos{\phi})}{4\varepsilon^2\sin^2\tilde\chi+\left[(\varepsilon{-}\eta^{+})\cos{\phi}{+}2(1{-}\varepsilon)\cos\tilde\chi\right]^2},
\end{align}
with $\tilde\chi=\chi_a+\chi_b$, one sees two contributions to the interference pattern. The transmission oscillates both via $\phi$, which is tuned with the magnetic flux, and via the kinetic phases $\chi=\chi_\alpha$, which depend on the circumference of the ring and on energy and therefore can be controlled by the potential $U_r$, as shown in Fig.~\ref{fig:2ttransmiss}. Note that in the symmetric case we are considering, ${\cal T}_{12}^{(2t)}(E)\propto1+\cos\phi$, hence total destructive interference occurs at odd multiples of $\phi=\pi$, irrespective of $\mu$. In less symmetric configurations, this condition depends on $\chi_a$ and $\chi_b$.

We distinguish two kinds of diamond-shaped regions in the $\phi$-dependence of the transmission, separated by resonances at $E=\epsilon_n$, cf. Fig.~\ref{fig:2ttransmiss}(a): around odd multiples of $\pi$, labeled as A in Fig.~\ref{fig:2ttransmiss}(a), destructive interference suppresses the transmission probability, which is exactly cancelled at $\phi=(2n+1)\pi$. The resonances hence do not cross each other at the crossings of the eigenstates there. Differently, states crossing at even multiples of $2\pi$ interfere constructively. In those regions, labelled as B in Fig.~\ref{fig:2ttransmiss}(a), the transmission is not totally suppressed between peaks. Note that the shape of the resonances depends on the magnetic flux~\cite{buttiker:1984}, see Fig.~\ref{fig:2ttransmiss}(b). This property has consequences for the thermoelectric performance, as discussed below. At energies corresponding to a single eigenmode, asymmetric single resonances appear, which can be used as filters for efficient heat to charge conversion~\cite{mahan:1996}. Differently, close to the crossing of two eigenmodes double-peaked resonances develop, with sharper tails, which are promising candidates for power improving thermoelectrics~\cite{whitney_most_2014}, see also Ref.~\onlinecite{behera_quantum_2023}. The dependence on energy is also interesting as the quadratic dependence in Eq.~\eqref{eq:eigenen} increases the separation between resonances with increasing $n$. This is to be taken into account for the thermoelectric response, as it typically increases when broken electron-hole symmetry is resolved within the range of thermal fluctuations. Hence, depending on the scale $\kBT$, the ring potential can be adjusted to enhance the response: higher temperatures require high enough $\mu-U_r$.

\begin{figure}[t]
    \includegraphics[width=\linewidth]{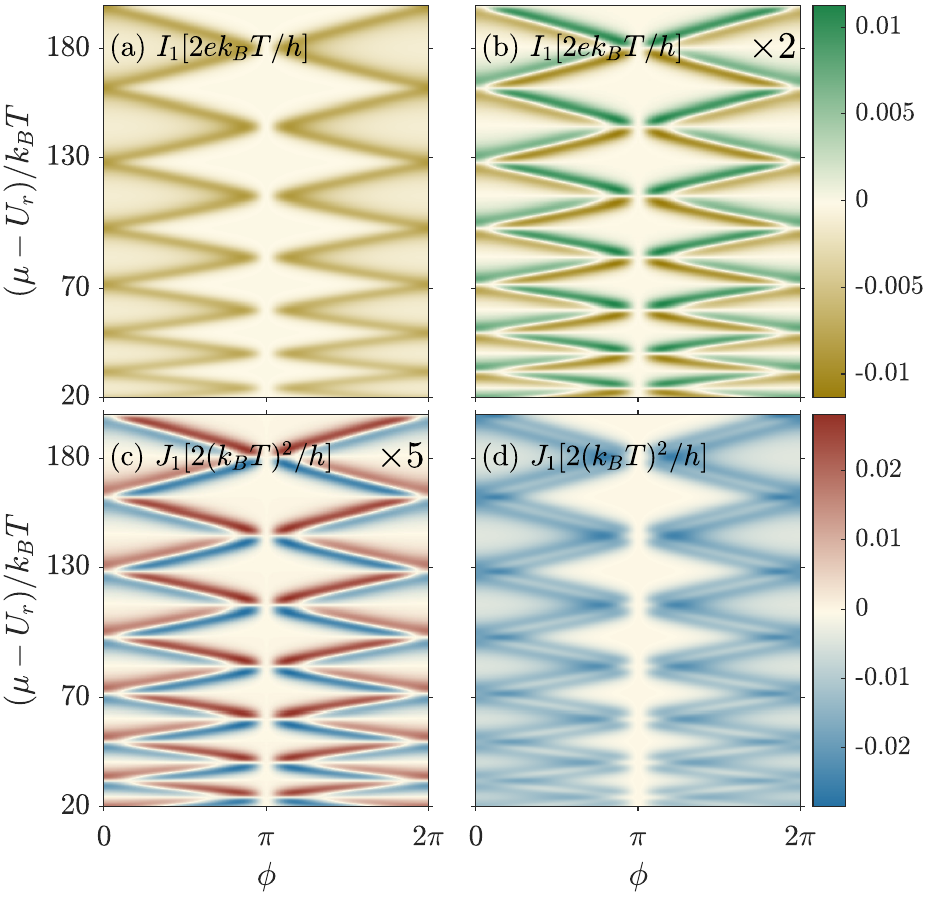}
    \caption{\label{fig:2tcurrents}\small Charge, $I$, and heat, $J$, currents as functions of the magnetic phase and the chemical potential for $L=l_0/2$ and $\varepsilon=0.25$, with (a),(c) $\Dmu=0.01\thinspace k_{\rm B}T$ and $\Delta T=0$, and (b),(d) $\Delta T/T=0.01\thinspace$ and $\Dmu=0$. }
\end{figure}
Charge and heat currents are shown in Fig.~\ref{fig:2tcurrents} for finite and small chemical potential and temperature differences ($T_1=T$, $T_2=T+\Delta T$,  $\mu_1=\mu+\Delta \mu/2$, $\mu_2=\mu-\Delta \mu/2$) for a ring of circumference $L=l_0/2$. Following the transmission in Fig.~\ref{fig:2ttransmiss}, currents are $2\pi$-periodic in $\phi$ and show diamond-shaped regions around $\phi=\pi$ where transport is suppressed by destructive interference. Notably, at $\phi=\pi$, both charge and heat currents are exactly cancelled, as ${\cal T}_{12}^{(2t)}(E)=0$ for all energies. As for the transmissions, the current peaks tend to separate with $\mu-U_r$. As we are considering relatively low temperatures, all resonances are clearly resolved. 

As can be seen in Figs.~\ref{fig:2tcurrents}(b) and \ref{fig:2tcurrents}(c), the sharp spectral features of the transmission induce a thermoelectric response. This manifests in the form of a Seebeck heat to charge converter when the two terminals support a temperature difference $\Delta T$, cf. Fig.~\ref{fig:2tcurrents}(b), and of Peltier cooling in the presence of a small electrochemical potential difference, $\Delta\mu$, cf. Fig.~\ref{fig:2tcurrents}(c). In both cases, the current changes sign around the transmission peaks, as expected for sharp resonances as those occurring in e.g., quantum dots~\cite{staring_coulomb_1993,dzurak_observation_1993,dzurak_thermoelectric_1997,scheibner_sequential_2007,svensson_lineshape_2012,svensson_nonlinear_2013,josefsson_quantum_2018}. As we are considering small differences within the linear regime, $\Delta\mu,\Delta T\ll\kBT$, the two effects show similar features, as corresponds from the Onsager reciprocity relations~\cite{butcher:1990}. Note that in a two terminal geometry, linear transport coefficients are even functions of the magnetic field, see also Eq.~\eqref{eq:T_two_terminals}.

\begin{figure}[t]
\includegraphics[width=\linewidth]{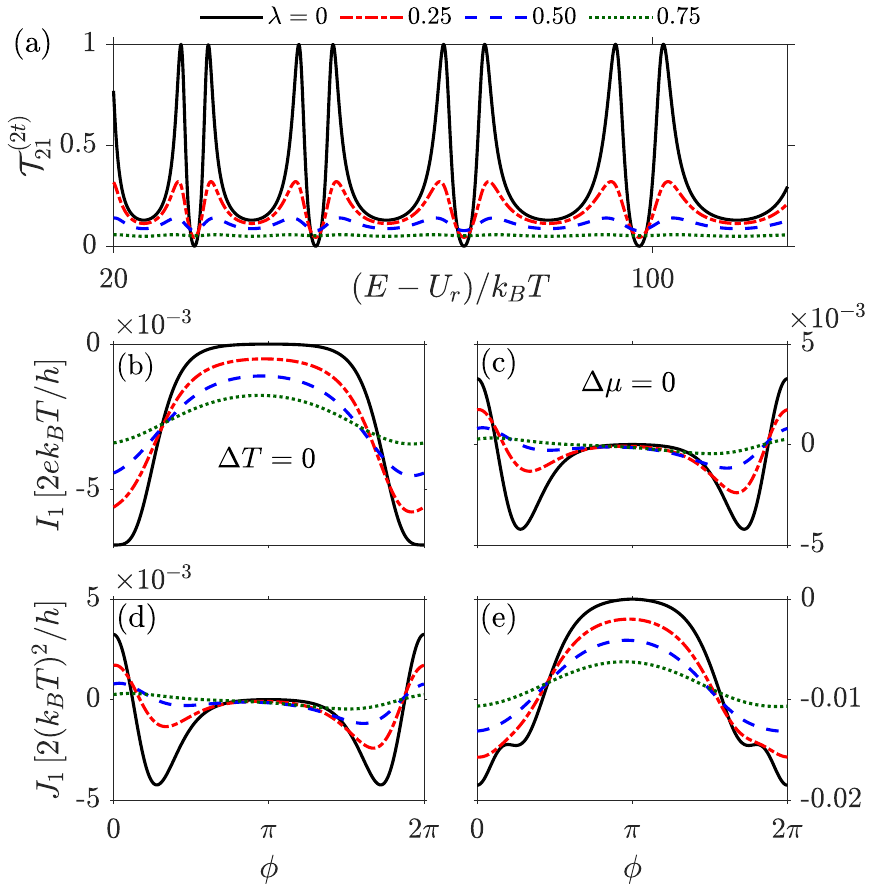}
\caption{\label{fig:2tdeph}\small Effect of dephasing. (a) Two-terminal transmission probability as a function of energy for $\mu-U_r = 100 \kBT$ and $\phi=\pi/4$, and (b),(c) charge and (d),(e) heat currents as functions of the magnetic flux, for different couplings to the dephasing probes, with (b),(d) $\Delta T=0$ and $\Delta\mu=0.01$ and (c),(e) $\Delta T/T=0.01$ and $\Delta\mu=0$. The magnetic field dependent oscillations disappear with the coupling to the probes, as expected. Other parameters: $\varepsilon=0.25$, $L=l_0/2$ 
}
\end{figure}

Let us now consider the effect of dephasing by coupling each segment to a dephasing probe with equal strength $\lambda_\alpha=\lambda$. By increasing the coupling, additional trajectories contribute where the electrons lose their phase before being absorbed by one of the transport terminals (1 and 2). These are not affected by interference, hence reducing the resonances, cf. Fig.~\ref{fig:2tdeph}(a). As dephasing increases, the role of the kinetic and magnetic phases is suppressed and the oscillations get increasingly flattened. In the strong dephasing limit $\lambda=1$, all injected electrons are absorbed by the dephasing probes and reinjected with a randomized phase (hence ${\cal T}_{21}^{(2t)}(E)=0$) so transport is fully incoherent. The impact on the currents is shown in Figs.~\ref{fig:2tdeph}(b)-(e). The flux dependence of the charge and thermal conductances is washed out and stays finite, as expected from an incoherent conductor, see Figs.~\ref{fig:2tdeph}(b) and (e). The thermoelectric response is however suppressed for any value of the flux, see Fig.~\ref{fig:2tdeph}(b), confirming that the thermoelectric response relies on the broken electron-hole symmetry induced by the kinetic phases~\cite{extrinsic}. 

\subsection{Thermoelectric generator}
\label{sec:2ttherm}

\begin{figure}[t]
    \includegraphics[width=\linewidth]{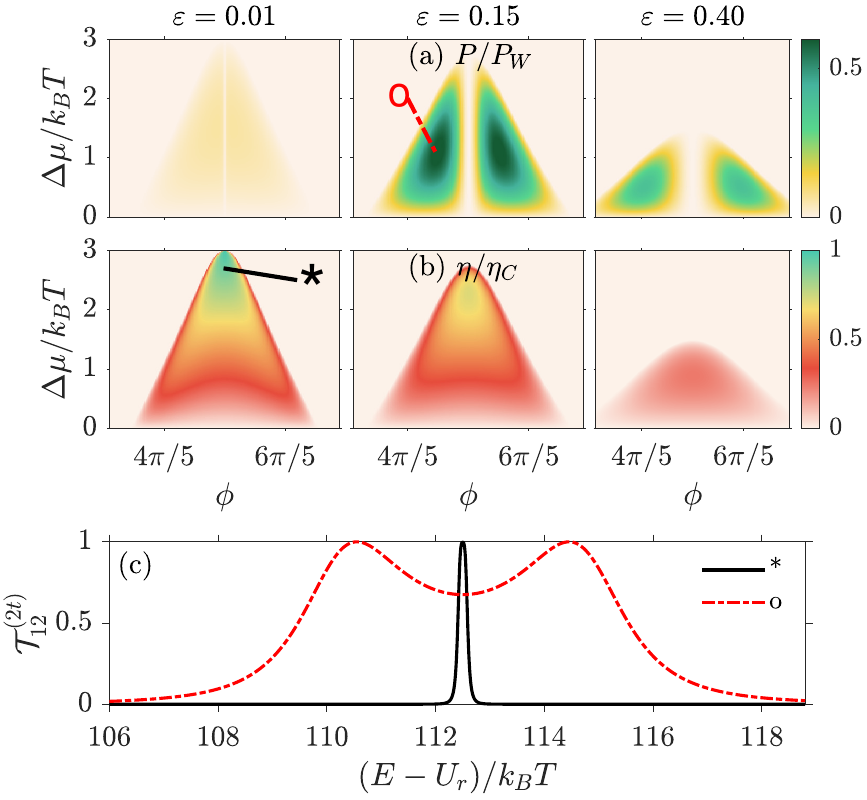}
    \caption{\label{fig:2t_Peff}\small (a) Power and (b) efficiency for $\varepsilon=0.01$ (left), 0.25 (center) and 0.40 (right column). Parameters: $L= l_{0}/2$, $\mu-U_r=108\thinspace k_{\rm B}T$, $\Delta T/T=1$. At low couplings, the low generated power is compensated by an efficiency close to Carnot's. While the efficiency decreases with $\varepsilon$, the power is maximal for intermediate couplings. (c) Transmission probability for the configurations corresponding to the maximal power $P_{max}$ (red dashed, $\phi=2.82$) and maximal efficiency $\eta_{max}$ (black full line, $\phi=3.132$) for the conditions indicated. }
\end{figure}
Let us evaluate the performance of the thermoelectric effect by the generated power and the efficiency $\eta=P/J_2$. We do this in Figs.~\ref{fig:2t_Peff}(a) and \ref{fig:2t_Peff}(b) for different couplings $\varepsilon$ and comparing to the thermodynamic bounds given by the Carnot efficiency $\eta_C=\Delta T/(T+\Delta T)$ and the Whitney-Pendry bound for the maximal power output from a single quantum channel~\cite{whitney_most_2014,pendry_quantum_1983}, $P_W=2A_0\pi^2\Delta T^2$, with $A_0=0.0321$. Weakly coupled rings generate small power outputs at a high efficiency~\cite{haack:2019} close to $\eta_C$. This occurs when a circulating state enters the transport window (defined by $\mu\pm2\kB(T+\Delta T)$), leading to a very narrow resonance which is known to result in high efficiencies~\cite{mahan:1996}. The width of the resonance depends on the coupling to the leads. Hence, increasing $\varepsilon$ affects the energy filtering and, as a consequence, the efficiency is reduced to a few tenths of $\eta_C$. 

The effect of coupling to the power is two-fold: 
on one hand, broad resonances allow for larger currents contributing to more power; on the other hand, energy resolution in filtering is lost, which harms the thermoelectric effect. A compromise is found at intermediate couplings where power is maximal, around $0.6P_W$, and efficiency is still around $\eta_C/2$, see Fig.~\ref{fig:2t_Peff}(a) and \ref{fig:2t_Peff}(b). The large power output can in this case also be attributed to the spectral properties of the ring: 
enabled by a sufficiently large $\varepsilon$, the overlap of two resonances occurring close to the crossing of the ring states result in a broad double peak whose tails are sharpened by the destructive interference, see Fig.~\ref{fig:2t_Peff}(c), a property that has been discussed to optimize the power output~\cite{whitney_most_2014,whitney_finding_2015,behera_quantum_2023,balduque_coherent_2023}. 
Remarkably, the position of the resonances is the same both for obtaining a maximal power and a maximal efficiency.

\begin{figure}[t]
    \includegraphics[width=\linewidth]{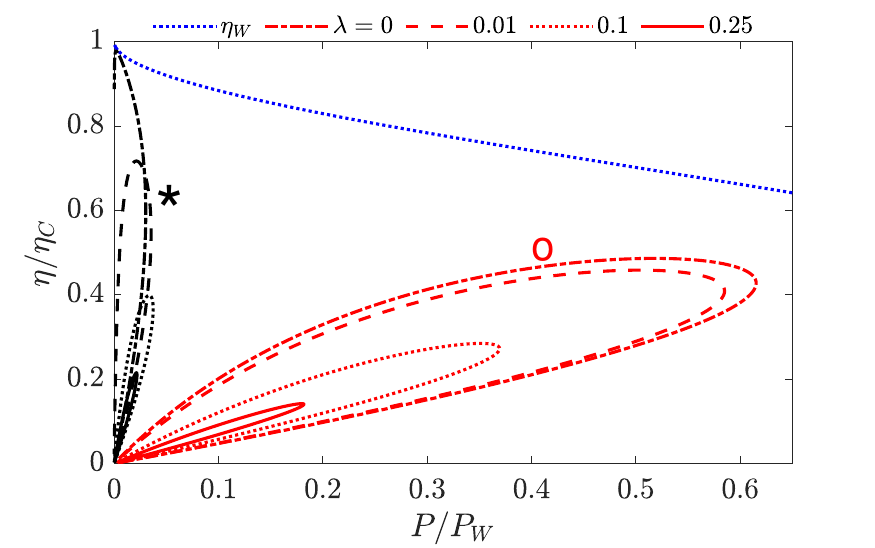}
    \caption{\label{fig:2t_lazo}\small Loop diagrams for efficiency vs. power for the optimal configurations considered in Fig.~\ref{fig:2t_Peff}(c) giving $P_{max}$ (red) and near $\eta_{max}$ (black). The effect of dephasing is shown by varying the coupling to the dephasing probes, $\lambda$. Parameters that are not indicated in the figure are as in Fig.~\ref{fig:2t_Peff}. 
    }
\end{figure}
The compromise between power and efficiency is further explored in Fig.~\ref{fig:2t_lazo}, which shows their mutual values as voltage is tuned from zero to the stall voltage where $I$ vanishes, and compared to the maximal value $P_{max}$ for a given efficiency obtained by a boxcar transmission~\cite{whitney_most_2014}. We do this in the two optimal configurations of Fig.~\ref{fig:2t_Peff}. It is found that the low coupling case approaches both bounds $\eta_C$ and $P_{max}$. However this result is very sensitive to dephasing. Optimizing power seems to be more robust: the case $\varepsilon=0.15$ maintains large power outputs (though far from the bound $P_{max}$) in the presence of dephasing, with efficiencies comparable to the case $\varepsilon=0.01$ for the same $\lambda$.

\subsection{All-thermal switch}
\label{sec:2tvalve}

\begin{figure}[t]
    \includegraphics[width=\linewidth]{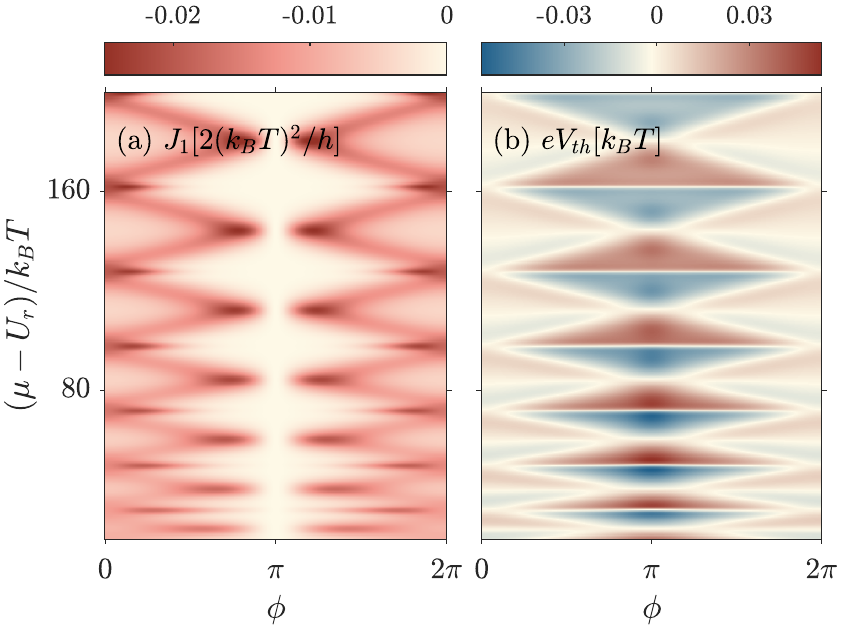}
    \caption{\label{fig:2t_JnoI}\small 
    (a) Heat current in the two-terminal all-thermal configuration  as a function of the tunable parameters: the magnetic flux and the ring potential. Tuning $\phi$ the system works as a thermal switch operating between a resonant transmission and a full cancellation of the current at $\phi=\pi$ for every value of $U_r$. Here, $\epsilon\thinspace =\thinspace 0.25$ and $L\thinspace =l_{0}/2$. (b) Thermovoltage $eV_{th}=\mu_2-\mu_1$ developed to satisfy the condition $I_l=0$~\cite{haack:2019}.
    }
\end{figure}
The transport properties shown in Fig.~\ref{fig:2tcurrents} can be used as a switching mechanism~\cite{haack:2019}: by simply sweeping the magnetic field, one changes the device from a resonantly conducting to an insulating state, both for particle and for heat currents. We are interested here in the later ones, and indeed assume the all-thermal condition where the charge current vanishes by considering an open conductor configuration. The two terminals then develop a chemical potential difference which coincides with the thermovoltage $\mu_2-\mu_1=eV_{th}$, plotted in Fig.~\ref{fig:2t_JnoI} with the resulting heat current. By comparing Figs.~\ref{fig:2tcurrents}(d) and \ref{fig:2t_JnoI}(a) one appreciates that the heat current is different is the short and open circuit configurations. In particular, double peaks in Figs.~\ref{fig:2tcurrents}(d) merge into wide single-featured resonances in Fig.~\ref{fig:2t_JnoI}(a) as the potential is tuned, and the double-peak resonances around even multiples of $\phi=\pi$ become considerably wider and extend into the negative-interference regions in the open circuit case. Of course, also in open circuit all currents vanish at $\phi=\pi$. As shown in Fig.~\ref{fig:2t_JnoI}(b), the developed chemical potential difference gives a measure of the thermoelectric origin of the heat current suppression. 

\begin{figure}[t]
\includegraphics[width=\linewidth]{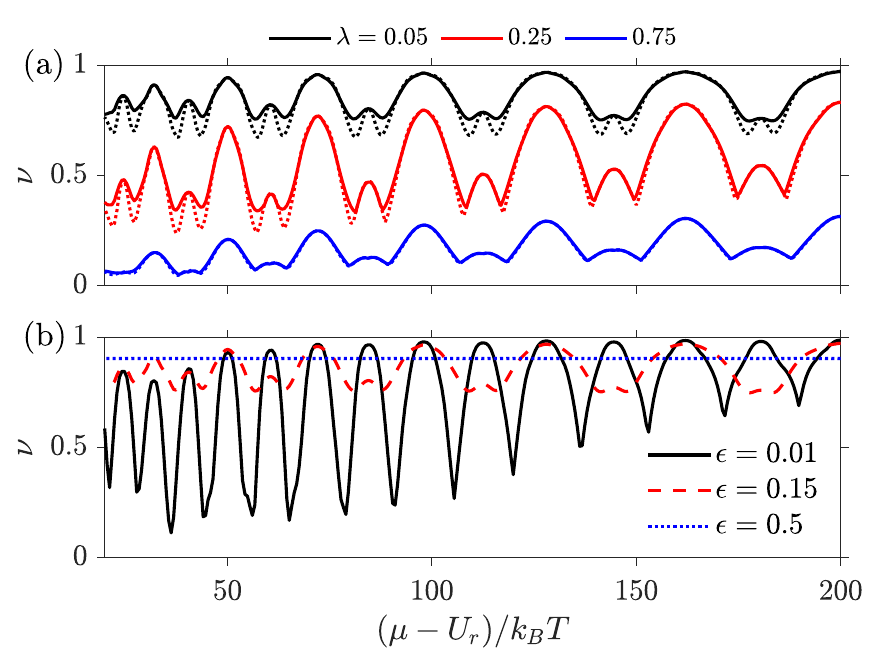}
\caption{\label{fig:2tswitch}\small Two-terminal switch coefficient as a function of the ring potential (a) for $\varepsilon=0.25$ and  $L=l_{0}/2$, for increasing coupling to the dephasing probes, $\lambda=\lambda_\alpha$, and (b) for $\lambda=0.05$ and different $\varepsilon$, with otherwise similar parameters ($\Delta T/T=0.01$, $\Delta \mu=0$).  
}
\end{figure}

We parameterize the switching operation by the visibility of the thermal interference pattern as a function of the magnetic flux for all other parameters fixed:
\beq
\nu(\phi)=\frac{J_{max}(\phi)-J_{min}(\phi)}{J_{max}(\phi)+J_{min}(\phi)},
\label{eq:valvecoeff}
\eeq
which is 1 if the current is totally blocked. In the case $\lambda=0$, the full destructive interference leads to $\nu=1$ at any condition. For finite coupling to the dephasing probes, the switch coefficient becomes parameter-dependent and $\nu<1$, see Fig.~\ref{fig:2tswitch}(a). However,  for low $\lambda$ the visibility can be tuned to still approach 1, especially for low coupling to the leads $\varepsilon\ll1/2$ and increasing $\mu-U_r$, as shown in Fig.~\ref{fig:2tswitch}(b). 

In Fig.~\ref{fig:2tswitch}(a) we compare the all-thermal (solid) and electro-thermal (dotted lines) switching. The two cases have similar maxima, however, the all-thermal switching is more stable under the variation of the chemical potential. The two cases become indistinguishable as dephasing increases.

\section{Three terminals}
\label{sec:3t}

\begin{figure}[t]
    \includegraphics[width=\linewidth]{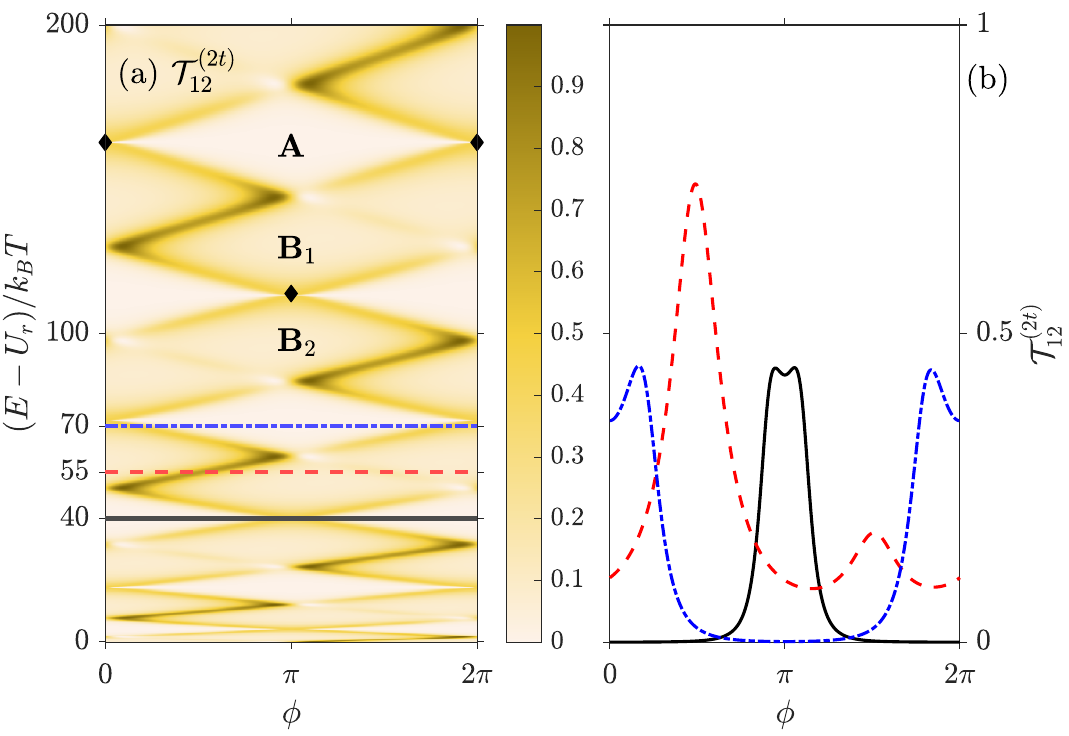}

    \caption{\label{fig:3ttransmiss}\small Three-terminal transmission probability between clockwise connected terminals, ${\cal T}_{12}(E,\phi)={\cal T}_{\circlearrowright}(E,\phi)={\cal T}_{\circlearrowleft}(E,-\phi)$, (a) as a function of energy and the magnetic phase, for a ring of circumference $L=l_0/2$ with  $\varepsilon=0.25$. (b) Cuts of the previous at the energies marked by the lines with the corresponding linestyles: $E-U_r=40\kBT, 55\kBT,70\kBT$. }
\end{figure}
We now extend the configuration to include a third terminal, as sketched in Fig.~\ref{fig:scheme}. We recall that we are assuming a symmetric configuration, with equidistantly separated junctions ($L_\alpha=L/3$) and equally coupled ($\varepsilon_l=\varepsilon$). This way, the necessary asymmetries involved in the proposed operations can only be attributed to the magnetic field induced asymmetry of the Onsager-Casimir reciprocity relations, ${\cal T}_{ij}(\Phi)={\cal T}_{ji}(-\Phi)$, which only play a role in multiterminal systems. Furthermore in our case, we can express it in terms of a single transmission coefficient, ${\cal T}_{\circlearrowright}(\Phi)={\cal T}_{12}(\phi)={\cal T}_{23}(\phi)={\cal T}_{31}(\phi)=|{\cal A}|^2$, with the transmission amplitude ${\cal A}={\cal N}/{\cal D}$ and 
\begin{align}
\label{eq:3t_ampl}
{\cal N}=&16i\varepsilon e^{i2\chi}\sin\chi\left[2e^{-i\chi}+e^{-i\phi}\eta^++2e^{i(\chi+\phi)}(\eta^+{-}1)\right]\nonumber\\
{\cal D}=&-16e^{-i\chi}+12e^{i\chi}(\eta^-)^2+4e^{i2\chi}\cos\phi(\eta^+)^3\\
&-3e^{i\chi}{\eta^-}^2[(\eta^-)^2-(\eta^+)^2]+e^{i5\chi}[(\eta^-)^2-(\eta^+)^2]^3/4,\nonumber
\end{align}
plotted in Fig.~\ref{fig:3ttransmiss}. Here again, $\chi=\chi_\alpha$.
The other transmission probabilities are ${\cal T}_{\circlearrowleft}(\phi)={\cal T}_{21}(\phi)={\cal T}_{32}(\phi)={\cal T}_{13}(\phi)$. From microreversibility we further get ${\cal T}_{\circlearrowleft}(\phi)={\cal T}_{\circlearrowright}(-\phi)$~\cite{jacquod_onsager_2012}. It is useful to write it in the weak coupling limit ($\varepsilon\rightarrow0$):
\beq
\label{eq:Tcirc_wc}
{\cal T}^w_{\circlearrowright}{=}2\varepsilon^2\sin^2{\chi}\frac{[1{+}\cos(2\chi)](1{+}\cos\phi){-}\sin(2\chi)\sin\phi}{\sin^2(3\chi)+64\cos^2\phi},
\eeq
 which gives a magnetoasymmetry $\delta{\cal T}={\cal T}_{\circlearrowright}-{\cal T}_{\circlearrowleft}\propto\sin\phi$ that changes sign around even multiples of $\phi=\pi$. Also ${\cal T}^w_{\circlearrowright}(\phi=(2n+1)\pi)=0$. This is enough to break inversion symmetry when heat is injected from one terminal, so it will flow preferentially into one of the other terminals, resulting on a circulator. It can also be used to induce a directed charge current between the other terminals, resulting in an energy harvester. Both effects are discussed below. 
In the opposite (strongly coupled) limit, $\varepsilon=1/2$, we get 
\beq
\label{eq:Tcirc_sc}
{\cal T}_{\circlearrowright}^s{=}16\sin^2(2\chi)\left[16+\cos(\chi-\phi)\right]/D ,
\eeq
with the denominator $D$ being an even function of $\phi$. Also in this case, $\delta{\cal T}\propto\sin\phi$.

\subsection{Heat currents}
\label{sec:3t-currents}

\begin{figure}[t]
    \includegraphics[width=\linewidth]{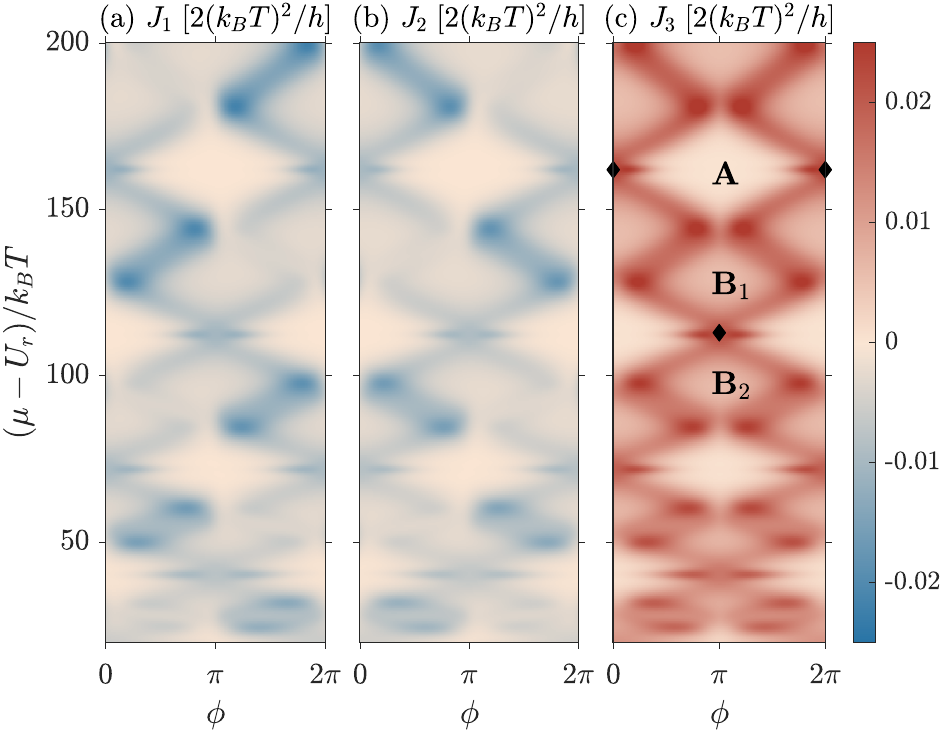}
    \caption{\label{fig:3t-J123}\small Heat currents as functions of the magnetic phase and the ring potential for the case where all terminals inject no charge, $I_l=0$, when terminal 3 is hot. Parameters: $\epsilon=0.25$, $L=l_0/2$, $\Delta T/T=0.01$.}
\end{figure}
In the discussion below, terminal 3 will serve as a heat source for being at an increased temperature, $T_3=T+\Delta T$, while terminals 1 and 2 remain at temperature $T$. The  heat current $J_3$ is split into terminals 1 and 2, as shown in Fig.~\ref{fig:3t-J123}, where we have imposed that all reservoirs are voltage probes, such that their chemical potentials satisfy $I_l=0$ (open circuit). 

\begin{figure}[t]
\includegraphics[width=\linewidth]{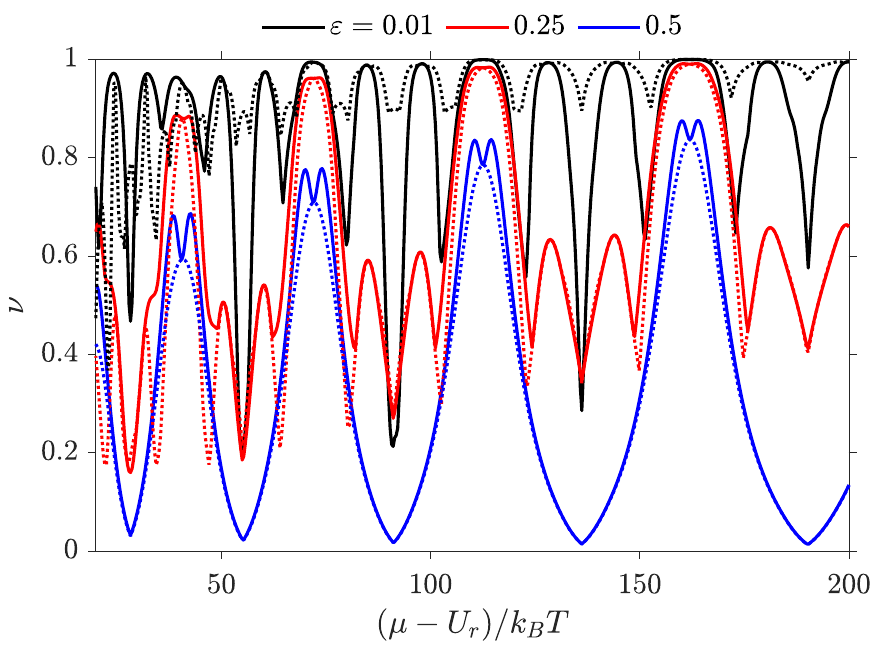}
\caption{\label{fig:3t_valve_Ur_L_allthermal}\small Three-terminal heat switch effect as a function of the ring potential with $L=l_0/2$ for different couplings. Full lines: all-thermal, dotted: electrothermal (with $\Delta \mu=0$). Parameters: $\Delta T/T=0.01$, $\lambda=0$. }
\end{figure}
We readily see that destructive interference in the ring yields regions where all currents vanish. Therefore the switch effect is also present in a three-terminal geometry. However, differently from the two-terminal case, the perfect suppression is limited by the interplay of the Aharonov-Bohm and the kinetic phases, cf. Eq.~\eqref{eq:3t_ampl}, so it does not occur for every electron density, in general. 
Interestingly, we find a regular sequence of three diamond-shaped regions centered around $\phi=\pi$, one of them insulating, labeled as A in Figs.~\ref{fig:3ttransmiss}(a) and ~\ref{fig:3t-J123}(c),  and the other two conducting, labeled as B$_1$ and B$_2$. 
We attribute this periodicity to the phase accumulated between terminals 1 and 2 being one third of the total phase $\sum_\alpha\chi_\alpha$ accumulated around the loop, which avoids the perfectly destructive interference except for some conditions determined by the ring potential (e.g., in the center of one every three diamonds)\footnote{We have confirmed that all diamonds are insulating for $J_3$ when $L_b=0$, and that sequences of $M$ diamonds with $M-1$ being conducting appear when $L_b=L/M$, in both cases with and $L_a=L_c=(L-L_b)/2$.}. Enhanced injection spots occur in the connection of two conducting regions, indicated with diamond shaped markers in Figs.~\ref{fig:3ttransmiss}(a) and~\ref{fig:3t-J123}(c), both at even and at odd multiples of $\phi=\pi$. At these spots, the current is equally absorbed by the other two terminals.

We show the switching parameter $\nu$ of $J_3$ as a function of the ring potential for different couplings in Fig.~\ref{fig:3t_valve_Ur_L_allthermal}. We find a doubly periodic behaviour predicted for non-negligible couplings which is less evident in the weak and strongly coupled regimes. In the tunneling regime with $\varepsilon\ll1$, the switch is most effective and tends to $\nu=1$. Optimal values are also obtained for intermediate couplings for sufficiently high $\mu-U_r$. 

We get an analytical understanding by comparing this behaviour with the electrothermal configuration (with all terminals having the same chemical potential, $\mu$, dotted lines in Fig.~\ref{fig:3t_valve_Ur_L_allthermal}), which shows similar properties in most  cases, but is less affected by the $\chi$-dependent oscillations at small $\varepsilon$. 
As terminals 1 and 2 are in the same equilibrium state (with the same temperature and chemical potential), described with the Fermi function $f_{eq}$, the integrand of $J_3$ is proportional to the spectral current $i_3=(2/h)[{\cal T}_{\circlearrowright}(\phi)+{\cal T}_{\circlearrowleft}(-\phi)](f_3-f_{eq})$.
In the limit $\varepsilon\rightarrow0$, we use Eq.~\eqref{eq:Tcirc_wc} to find $i_3\propto\sin^2\chi[1+\cos(2\chi)](1+\cos\phi)$, i.e. it vanishes at $\phi=\pi$, resulting in $\nu\rightarrow1$, in agreement with Fig.~\ref{fig:3t_valve_Ur_L_allthermal}. 
In the strongly coupled ring, $\varepsilon\rightarrow1/2$, from Eq.~\eqref{eq:Tcirc_sc} we find $i_3\propto\sin^2\chi(16+\cos\chi\cos\phi)$, which describes the $\pi$-periodic in $\chi$ vanishing of $\nu$. In this case, $\phi$ is not able to totally cancel the current, thus $\nu<1$. However, one can use the kinetic phase to define a switch by tuning $\mu-U_r$ for constant $\phi$. At intermediate couplings, the double periodic dependence can be explained by the injected electrons accumulating two different phases (either $\chi$ or $2\chi$) before they are absorbed.

\subsection{Circulator}
\label{sec:circ}

The finite magnetoasymmetry $\delta{\cal T}$ involves that ${\cal T}_{13}\neq{\cal T}_{23}$, i.e., electrons injected from terminal 3 will have a preferred terminal to be absorbed, despite they being in a similar thermodynamic state and the conductor being perfectly symmetric (spatially). One already sees in Fig.~\ref{fig:3t-J123} that $J_1(\phi)\neq J_2(\phi)$, but rather $J_1(\phi) = J_2(-\phi)$. We therefore define a circulation coefficient that measures this effect, considering that heat is injected from terminal 3, as 
\beq
{\cal C}_3=\frac{J_1-J_2}{J_3}.
\eeq
It is bounded by $\pm1$ when all heat injected by 3 is absorbed by a single terminal. However, this only occurs in chiral setups like quantum Hall edge states~\cite{chiraldiode,granger_observation_2009,nam_thermoelectric_2013}, as long as heat transport is avoided through the bulk. 

\begin{figure}[t]
\includegraphics[width=\linewidth]{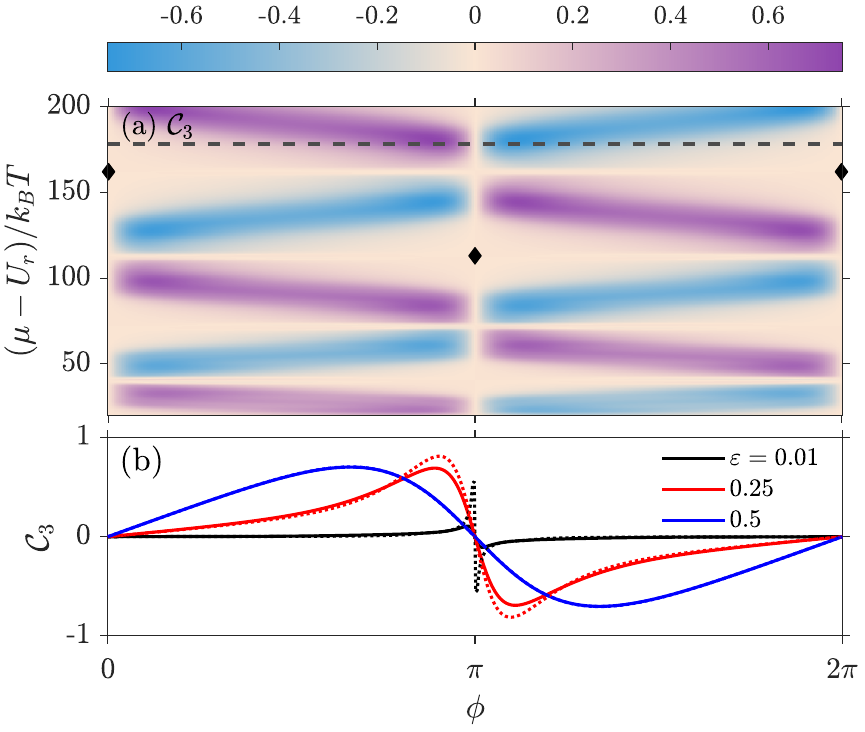}
\caption{\label{fig:circul_Ur_allthermal}\small (a) All-thermal circulation coefficient as a function of $\phi$ and the ring potential for $\varepsilon$=0.25 and $L=l_0/2$,  with  $\Delta T=0.01$. (b) Cuts for $(\mu-U_r)/k_BT=177$ (solid line in (a)) and various couplings for the all-thermal (full lines) and the electrothermal (dotted) configurations.} 
\end{figure}
The result is plotted in Fig.~\ref{fig:circul_Ur_allthermal}, where ${\cal C}_3$ changes sign around integer multiples of $\phi=\pi$, as expected from the Onsager-Casimir reciprocity relations, and around the potentials where the eigenstates cross between the two conducting diamonds B$_1$ and B$_2$, marked by symbols in Fig.~\ref{fig:circul_Ur_allthermal}(a) (see also Fig.~\ref{fig:3t-J123} and the discussion therein), forming a rhomboidal-shaped structure. This evidences that some of the circulating states do not contribute to ${\cal C}_3$. 

The circulator achieves coefficients close to 0.8 for intermediate values of the coupling. Both for strongly and very weakly coupled systems the performance is reduced. In particular, weakly coupled systems give a finite ${\cal C}_3$ only very close to $\phi=\pi$, see Fig.~\ref{fig:circul_Ur_allthermal}(b). This is expected, as in our case both ${\cal T}^w_{\circlearrowright}$ and ${\cal T}^w_{\circlearrowleft}$ vanish simultaneously at $\phi=\pi$, see Eq.~\eqref{eq:Tcirc_wc}. Asymmetric rings will not have such constriction and are expected to improve ${\cal C}_3$.  

We compare the all-thermal and electrothermal cases in Fig.~\ref{fig:circul_Ur_allthermal}(b). As for the thermal switch, they have very similar behaviour, with the electrothermal device outperforming the all-thermal one, specially at low couplings. In the electrothermal condition, we can again get analytical insight, since 
\beq
J_1-J_2=\frac{2}{h}\int_{-U_r}^\infty{dE} (E-\mu)\delta{\cal T}(E)[f_{eq}(E)-f_{3}(E)],
\eeq
i.e., $J_1-J_2\propto\sin\phi$. The sign of the circulation depends however also on the energy dependence of the kinetic phase, which needs to be integrated.

\subsection{Nonlocal thermoelectric engine}
\label{sec:engine}

\begin{figure}[t]
\includegraphics[width=\linewidth]{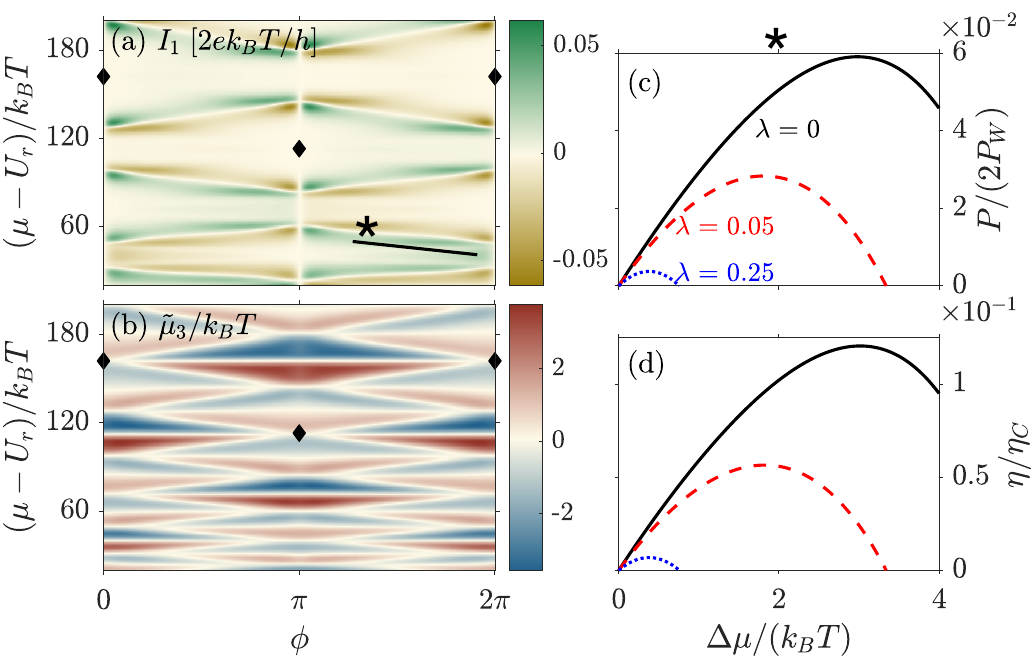}
\caption{\label{fig:3Theatengine}\small (a) Generated current for $\Delta\mu=0$, and (b) developed potential in the hot source, as a function of the magnetic flux and the internal potential in the nonlocal heat engine configuration, when reservoir 3 is at a higher temperature with $\Delta T/T=1$. (c) Power and (d) efficiency for the configuration marked by $\star$ (at $\phi=6$, $\mu-U_r=41\kBT$) in (a), as functions of the applied voltage and in the presence of dephasing. Parameters: $\varepsilon=0.15$, $L=l_0/2$}
\end{figure}
The three-terminal setup can also be used as an energy harvester: terminal 3 is coupled to a heat source that injects a current $J_3$ with $I_3=0$. Previous works have found that this heat currents are able to induce a charge current between the other two terminals (having the same temperature $T$ and chemical potential $\mu$) provided the conductor has broken electron-hole and inversion symmetries~\cite{sothmann:2015,wang_inelastic_2022}. In coherent conductors, the electron-hole symmetry is broken by the kinetic phase~\cite{hofer:2015,Vannucci2015, samuelsson:2017,haack:2019,haack_nonlinear_2021,genevieveqpc,extrinsic}. Here we propose the magnetic phase to also break the inversion symmetry. The argument is simple: with $f_1=f_2=f_{eq}$, the elastic transport between 1 and 2 vanishes, and the charge current in the hot terminal is
\beq
I_3={-}I_1{-}I_2=\frac{2e}{\hbar}\int{dE}[{\cal T}_{\circlearrowright}(E){+}{\cal T}_{\circlearrowleft}(E)][f_3(E){-}f_{eq}(E)],
\eeq
with the two terms within the first pair of brackets being the contribution of the current absorbed by terminals 1 and 2, respectively.  
Thus, in order to have $I_3=0$, the magnetoasymmetry makes it possible that a $\mu_3=\tilde\mu_3$ develops so that these two contributions have equal magnitude but are opposite in sign. We write the distribution of terminal 3 at that chemical potential as $\tilde{f}_3$. Hence
\beq
I_1=-I_2=\frac{2}{\hbar}\int{dE}{\cal T}_{\circlearrowright}(E)[f_{eq}(E){-}\tilde{f}_3(E)]
\eeq
is the generated current that is measured between 1 and 2 (with the sign defined by the details of the transmission) via the inelastic scattering at terminal 3. The result is plotted in Fig.~\ref{fig:3Theatengine}(a). 

We get an intuition of the resulting current by performing a Sommerfeld expansion (valid in the linear regime at temperatures small compared with the range of variation of the transmission probabilities~\cite{extrinsic}):  
\beq
\frac{I_1}{2q_H}=\frac{{\cal T}_{\circlearrowright}(\mu,\phi){\cal T}'_{\circlearrowright}(\mu,{-}\phi){-}{\cal T}_{\circlearrowright}(\mu,{-}\phi){\cal T}'_{\circlearrowright}(\mu,\phi)}{{\cal T}_{\circlearrowright}(\mu,\phi)+{\cal T}_{\circlearrowright}(\mu,-\phi)}\Delta T
\eeq
in terms of a single transmission coefficient and its energy derivative, ${\cal T}'_{\circlearrowright}=\partial_E{\cal T}_{\circlearrowright}$, where $q_H=\pi^2k_{\rm B}^2T/3h$. The chemical potential developed in the hot reservoir is 
\beq
\tilde{\mu}_3=\mu-h q_H\frac{{\cal T}'_{\circlearrowright}(\mu,\phi)+{\cal T}'_{\circlearrowright}(\mu,{-}\phi)}{{\cal T}_{\circlearrowright}(\mu,\phi)+{\cal T}_{\circlearrowright}(\mu,-\phi)}\Delta T,
\eeq
Note that in our symmetric configuration, at the crossing points between conducting regions [marked by diamond-shaped symbols in Fig.~\ref{fig:3ttransmiss}(a)], one can further simplify this expression by noticing that ${\cal T}_{\circlearrowright}(\mu,-\phi)={\cal T}_{\circlearrowright}(\mu,\phi)$, see Fig.~\ref{fig:3ttransmiss}(b), hence 

\beq
\label{eq: analytical_I1}
I_1=-q_H\partial_E(\delta{\cal T})
\eeq
even if $\delta\mu_3=0$ at those conditions, see Fig.~\ref{fig:3Theatengine}(b). In other words, the generated current gives a direct measure of the Onsager-Casimir reciprocity relations. 

The highest values for the currents coincide with the regions where ${\cal C}_3$ is maximal (at the outer edges of the B$_1$ and B$_2$ regions, see Figs.~\ref{fig:3ttransmiss}(a) and \ref{fig:circul_Ur_allthermal}(a)). These features emphasize the role of the magnetoasymmetry in enabling the thermoelectric effect. 

Although smaller, a finite current is also generated within the insulating regions A, see for example the region marked by a star symbol. This is a consequence of having a finite temperature, which allows the nearby resonances for energies above $\mu-U_r$ for ${\cal T}_{\circlearrowright}(\mu,\phi)$ and below $\mu-U_r$ for ${\cal T}_{\circlearrowleft}(\mu,\phi)$ to contribute to transport, with their derivatives contributing with the same sign to the current, cf. Eq.~\eqref{eq: analytical_I1}. This property, identified as the optimal condition for energy harvesting in gate-controlled devices~\cite{jordan:2013,whitney_quantum_2016,balduque_coherent_2023} is here uniquely due to magnetic reciprocity relations. 
While other configurations close to $\phi=\pi$ show a larger current at $\Delta\mu=0$ (due to narrow resonances), the cooperation of reciprocal transport coefficients at the condition marked by a star in Fig.~\ref{fig:3Theatengine}(a) provides a much larger power and efficiency ($\eta=P/J_3$) at finite $\Delta\mu$, as plotted in Figs.~\ref{fig:3Theatengine}(c) and \ref{fig:3Theatengine}(d).

Both the generated power and the efficiency are modest in absolute terms, see Figs.~\ref{fig:3Theatengine}(c) and \ref{fig:3Theatengine}(d). However, comparing with previous interference-induced nonlocal engines with no explicit energy-dependent scattering regions~\cite{extrinsic}, the nonreciprocal effect leads to a considerably enhanced thermoelectric performance. This can always be optimized by considering energy-dependent junctions with $\varepsilon(E)$ (e.g., including resonant-tunneling barriers~\cite{balduque_coherent_2023}) or allowing for asymmetric configurations.

\section{Conclusions}
\label{sec:conclusion}

We have proposed the magnetic flux through annular junctions of coherent conductors to control the thermal and thermoelectric effects via quantum interference in multiterminal devices. We consider a minimal model of a single-channel ring connected either to two or three terminals, finding that the interplay of kinetic and magnetic phases displays a rich interference pattern with sharp resonances due to the presence of circulating currents, and regions with destructive interference which can be tuned both with external gates and magnetic fields. We provide an interpretation of these features in terms of the spectral properties of the junction, in particular the symmetries of its transmission probabilities. 

In two-terminal configurations, the system can be used as a thermoelectric generator achieving either high efficiencies or high output power when using either very sharp peaks (in the weak coupling limit) or wider double resonances (in intermediate coupling regimes). Conditions of destructive interference which do not depend on energetic details induce a magnetic thermal switch whose efficiency remains high in the presence of dephasing. 

We introduce the three-terminal geometry which allows for more complex functionalities when heat is injected from the third terminal. We investigate all-thermal and electro-thermal configurations. The spatial separation of the cold-terminal junctions introduces additional phases that avoid the thermal switch effect to occur for every ring potential. Nevertheless, efficient switching for intermediately coupled rings can be found for wide regions of the ring potential. The magnetic field induced nonreciprocity of the transport coefficients can be used to define a thermal circulator able to perform efficiently even in the strongly coupled regime. 

The kinetic and magnetic phases are sufficient to break the electron-hole and inversion symmetries, as needed for a conductor to work as a nonlocal thermoelectric engine. We find that the Aharonov-Bohm effect enhances the  performance both in terms of the output power and the efficiency, compared to three-terminal analogues using only kinetic phases~\cite{extrinsic}, with optimal configurations enabled by reciprocity relations. 

We have considered a symmetric configuration which emphasizes the role of quantum phases in symmetry breaking. In realistic configurations, asymmetries in the distance between junctions may compromise the occurrence of destructive interference, which can however be controlled by additional gate potentials. Exploring these asymmetries and allowing for explicitly energy-dependent scatterers may also be a way to further increase the performance of the system as a thermal circulator and converter. 

%

\begin{acknowledgments}
We thank G\'eraldine Haack for comments on our manuscript. We acknowledge funding from the Spanish Ministerio de Ciencia e Innovaci\'on via grants No. PID2019-110125GB-I00 and No. PID2022-142911NB-I00, and through the ``Mar\'{i}a de Maeztu'' Programme for Units of Excellence in R{\&}D CEX2023-001316-M.
\end{acknowledgments}

\bibliography{biblio.bib}

\end{document}